\documentclass[aps,prc,floatfix,groupedaddress,showpacs,amsfonts,nofootinbib,twocolumn]{revtex4}
\usepackage{bm}
\usepackage{epsfig}
\usepackage{amsmath}
\usepackage{dcolumn}

\begin{document}

\title{Higher-order corrections to electron-scattering multipoles}

\author{J. Grineviciute \footnote{Present address: Physique Nucl{\'e}aire Th{\'e}orique et Physique Math{\'e}matique, C.P. 229, Universit{\'e} Libre de Bruxelles (ULB), B 1050 Brussels, Belgium}}
\affiliation{Department of Physics, Western Michigan University, Kalamazoo, MI 49008}

\author{Dean Halderson}
\affiliation{Department of Physics, Western Michigan University, Kalamazoo, MI 49008}

\begin{abstract}
A procedure is suggested for calculating electro-excitation multipoles to order $1/M^{2}_{N}$ with only the operators required in the calculations to order $1/M_{N}$.  It is also shown that calculations to order $1/M^{2}_{N}$ cannot account for the contributions of a fully relativistic calculation of the transverse response.
\end{abstract}


\maketitle

The role of relativity in ($e,e^{\prime}p$) has been investigated by many authors \cite{HJ95,U95,U99,U98,JO94}.  In most of these works, a Dirac equation with vector and scalar potentials is written for the outgoing proton, and then various elements of the Dirac equation are investigated as one proceeds to make a non-relativistic reduction.  For example in Ref. \cite{U95}, a Darwin factor, containing the potentials, was shown to reduce the upper component of the outgoing proton wave function in the interior, and hence, produce lower ($e,e^{\prime}p$) cross sections than the equivalent non-relativistic calculation.  In a similar fashion this potential containing term was included in the nuclear current in Ref. \cite{HJ95}.  This Brief Report is concerned with including relativistic effects in non-relativistic calculations for ($e,e^{\prime}$) processes where the vector and scalar potentials are unavailable. 

The standard procedure for describing electron-nucleus scattering is to assume a current-current interaction.  In a Born approximation for the electron, the electron current takes the simple form of a free Dirac particle.  The nuclear current would ideally come from a relativistic many-body calculation.  However, sophisticated structure calculations for light systems are basically non-relativistic.  The purpose of this brief report is two-fold.  The first is to provide a suggestion for including relativistic effects to order $1/M^{2}_{N}$ and second, to demonstrate that this procedure, and related procedures, can be inaccurate when describing transverse responses.   

Two approaches have been employed to derive the second order corrections to the nuclear current.   In Ref. \cite{MV62} the authors begin with a single particle Dirac equation for a spin-1/2 particle in a given electromagnetic field \cite{BG80},
\begin{equation}
\label{Eq1}
\left(i\gamma^{\mu}\partial_{\mu}-eF_{1}\gamma^{\mu}A_{\mu}+\frac{e}{4M}KF_{2}\sigma_{\mu\nu}F^{\mu\nu}-M \right)\psi =0  ,
\end{equation}
where $F^{\mu\nu}=\partial^{\nu}A^{\mu}-\partial^{\mu}A^{\nu}$, and $A^{\mu}$ is taken to be the M{\o}ller potential \cite{U98} of the electron,
\begin{equation}
\label{Eq2}
A^{\mu} = \frac{m_{e}e}{q^{2}}\frac{1}{\left(E_{0}E^{\prime}_{0}\right)^{1/2}} e^{-iq_{\nu}x^{\nu}} \bar{u}^{\prime} \gamma u .
\end{equation}
A Foldy-Wouthuysen transformation decouples Eq. (\ref{Eq1}) into one which is non-relativistic from which the second-order corrections are obtained and one that describes the negative-energy states.  The advantage of this procedure is that it can also give higher order corrections.  The zeroth and first order multipoles from this procedure are
\begin{equation}
\label{Eq3}
M^{Coul}_{J\: M} = \sum_{i} j_{J}\left(qr_{i}\right) Y_{J\: M}\left(\hat{r}_{i}\right) F^{i}_{1}\left(q^{2}_{\mu}\right) ,
\end{equation}
{\small\begin{align}
\label{Eq4}
T^{el}_{J\: M} & = \sum_{i} \left(F^{i}_{1}\left(q^{2}_{\mu}\right) / M_{N}\right) \nonumber\\
& \times  \left\{-\left(\frac{J}{2J+1}\right)^{1/2} j_{J+1}\left(qr_{i}\right) \left[Y_{J+1}\left(\hat{r}_{i}\right)\otimes\vec{\nabla}_{i} \right]_{J\: M}    \right.\nonumber\\
&+ \left. \left(\frac{J+1}{2J+1}\right)^{1/2} j_{J-1}\left(qr_{i}\right) \left[Y_{J-1}\left(\hat{r}_{i}\right)\otimes\vec{\nabla}_{i} \right]_{J\: M}     \right\}\nonumber\\
&+\left[F^{i}_{1}\left(q^{2}_{\mu}\right) +K_{i}F^{i}_{2}\left(q^{2}_{\mu}\right) \right] \left[q/\left(2M_{N}\right) \right] j_{J}\left(qr_{i}\right)  \nonumber\\
& \times \left[Y_{J}\left(\hat{r}_{i}\right)\otimes\sigma_{i}\right]_{J\: M}  ,
\end{align}}
{\small\begin{align}
\label{Eq5}
T^{mag}_{J\: M} &= \sum_{i} \left(iq\right) \left\{-\left(\frac{J}{2J+1}\right)^{1/2} \right. \nonumber\\
& \times j_{J+1}\left(qr_{i}\right) \left[Y_{J+1}\left(\hat{r}_{i}\right)\otimes \sigma_{i} \right]_{J\: M}    \nonumber\\
&+\left. \left(\frac{J+1}{2J+1}\right)^{1/2} j_{J-1}\left(qr_{i}\right) \left[Y_{J-1}\left(\hat{r}_{i}\right)\otimes\sigma_{i} \right]_{J\: M}     \right\}\nonumber\\
&+\left[F^{i}_{1}\left(q^{2}_{\mu}\right) +K_{i}F^{i}_{2}\left(q^{2}_{\mu}\right) \right] /\left(2M_{N}\right)  \nonumber\\
&-\left[iF^{i}_{1}\left(q^{2}_{\mu}\right)/M_{N} \right] j_{J}\left(qr_{i}\right) \left[Y_{J}\left(\hat{r}_{i}\right)\otimes\vec{\nabla}_{i}\right]_{J\: M} .
\end{align}}
With the replacements $F^{i}_{1}\left(q^{2}_{\mu}\right)=F^{i}_{2}\left(q^{2}_{\mu}\right)=1$ or 0 for $p$ and $n$ and $F^{i}_{1}\left(q^{2}_{\mu}\right) +K_{i}F^{i}_{2}\left(q^{2}_{\mu}\right)=\mu_{i}$, these expressions become the commonly employed multipoles of Ref. \cite{FW65}.  

A second approach starts with a covariant electromagnetic current density,
{\small\begin{equation}
\label{Eq6}
\hat{J}^{\mu}\left(x\right) = e_{i} \bar{\psi}_{f}\left(x\right) \gamma^{\mu} \psi_{i}\left(x\right)+\frac{e_{i}}{2M} \partial_{\mu}\left[ \bar{\psi}_{f}\left(x\right) K \sigma^{\mu\nu}  \psi_{i}\left(x\right)  \right] .
\end{equation}}
The functions $\psi_{i}$ and $\psi_{f}$ are assumed to be solutions of a single-particle Dirac equation with energies $E_{i}$ and $E_{f}$.  With the substitution for the time derivatives, 
\begin{equation}
\label{Eq7}
\frac{\partial}{\partial t}\left(\psi^{+}_{f}K\beta\vec{\alpha} \psi_{i} \right)=i\left[H,\left(\psi^{+}_{f}K\beta\vec{\alpha} \psi_{i} \right) \right] ,
\end{equation}
Ref. \cite{F85} was able to provide compact expressions for the electromagnetic multipole operators.  These expressions are repeated here in the notation of Ref. \cite{DH79}:
{\small\begin{equation} 
\label{Eq8}
 M^{Coul}_{JM} = 
\begin{pmatrix}
 {e_{i}j_{L}\left(qr\right)Y_{JM}\left(\hat{r}\right)} & {-iq \left(K_{i}/2M\right) \Sigma^{'' }_{J M} } \\
 {iq \left(K_{i}/2M\right) \Sigma^{'' }_{J M} } & {e_{i}j_{L}\left(qr\right)Y_{JM}\left(\hat{r}\right)} 
\end{pmatrix}
 \; ,
\end{equation}}
{\begin{widetext}\begin{equation}
\label{Eq9}
 T^{el}_{JM} = 
\begin{pmatrix}
 { q\left(K_{i}/2M\right)  \Sigma_{JM}} & { i \left[e_{i}+ K_{i}\left(E_{f}-E_{i}\right)/2M\right]   \Sigma^{\prime}_{JM} } \\
 {i \left[e_{i}- K_{i}\left(E_{f}-E_{i}\right)/2M\right]   \Sigma^{\prime}_{JM}  } & {-q\left(K_{i}/2M\right)  \Sigma_{JM}} 
\end{pmatrix}
\end{equation}
\begin{equation}
\label{Eq10} 
 T^{mag}_{JM} =
\begin{pmatrix}
 { i q\left(K_{i}/2M\right) \Sigma^{\prime}_{JM}} & {   \left[e_{i}+ K_{i}\left(E_{f}-E_{i}\right)/2M\right]  \Sigma_{JM}  } \\
 { \left[e_{i}- K_{i}\left(E_{f}-E_{i}\right)/2M\right]  \Sigma_{JM} } & {- i q\left(K_{i}/2M\right) \Sigma^{\prime}_{JM}} 
\end{pmatrix}
\; ,
\end{equation}
\end{widetext}}
with 
\begin{equation}
\label{Eq11}
\Sigma_{JM}=j_{J}\left(qr_{i}\right)\left[Y_{J}\left(\hat{r}_{i}\right)\otimes\sigma_{i} \right]_{JM} ,
\end{equation}
{\small\begin{align}
\label{Eq12}
\Sigma^{\prime}_{JM}=& \left\{ -\left(\frac{J}{2J+1}\right)^{1/2} j_{J+1}\left(qr_{i}\right) \left[Y_{J+1}\left(\hat{r}_{i}\right)\otimes \sigma_{i} \right]_{J M} \right. \nonumber\\
& +\left. \left(\frac{J+1}{2J+1}\right)^{1/2} j_{J-1}\left(qr_{i}\right) \left[Y_{J-1}\left(\hat{r}_{i}\right)\otimes \sigma_{i} \right]_{J M}  \right\} ,
\end{align}
\begin{align}
\label{Eq13}
\Sigma^{'' }_{J M} =& \left\{ \left(\frac{J+1}{2J+1}\right)^{1/2} j_{J+1}\left(qr_{i}\right) \left[Y_{J+1}\left(\hat{r}_{i}\right)\otimes \sigma_{i} \right]_{J M}\right. \nonumber\\
&\left. + \left(\frac{J}{2J+1}\right)^{1/2} j_{J-1}\left(qr_{i}\right) \left[Y_{J-1}\left(\hat{r}_{i}\right)\otimes \sigma_{i} \right]_{J M}  \right\} ,
\end{align}}

The operators act on wave functions of the form
\begin{equation} 
\label{Eq14}
\psi =
\begin{pmatrix}
{\left[F\left(r\right)/r\right]\Phi_{\kappa m}}\\
{\left[iG\left(r\right)/r\right]\Phi_{- \kappa m}}
\end{pmatrix}
=
\begin{pmatrix}
{\eta_{U}}\\
{\eta_{L}}
\end{pmatrix}
 ,
\end{equation}
where
\begin{equation}
\label{Eq15}
\Phi_{\kappa m} = \sum_{m_{l} m_{s}} {C^{l\ 1/2\ j}_{m_{l} m_{s} m}
Y_{l m_{l}}\left(\theta , \phi \right)\chi_{m_{s}} }   \; ,
\end{equation}
$j=\left|\kappa\right|-1/2$, and $l=\kappa$ for $\kappa>0$, but $l=-\left(\kappa+1\right)$ for $\kappa<0$.

One notes that these expressions require calculating the same operators that appear in Eqs. (\ref{Eq3})-(\ref{Eq5}).  However, these expressions also require knowledge of the lower component of the wave function which would be unavailable in non-relativistic shell model calculations.  An approximation can be employed at this point as in Ref. \cite{ODW72} by noting that the lower component of the single-particle Dirac equation with no potential is given by 
\begin{equation}
\label{Eq16}
G\rightarrow \tilde{G}=\frac{\vec{\sigma}\cdot \vec{p}}{E+M}\eta_{U}
\end{equation}
This substitution into Eqs. (\ref{Eq8})-(\ref{Eq10}) yields Eqs. (\ref{Eq3})-(\ref{Eq5}) plus terms of order $1/M^{2}_{N}$.  This approach has been referred to as ``direct Pauli reduction'' \cite{FP94}. Some of the second order terms differ from those derived from the Foldy-Wouthuysen transformation.  The two approaches have been discussed in Ref. \cite{FP94}, where it was concluded that the Foldy-Wouthuysen approach gives the correct answer for the second order S-matrix in some electromagnetic processes, and the direct Pauli reduction omits negative energy states in the intermediate states.  However, the authors state that the situation is ambiguous when the potentials are not known, as in the case of a non-relativistic shell model calculation, where the Dirac potentials are not known.  This work continues with the direct Pauli reduction because of the clarity of the approximation in Eq. (\ref{Eq16}).

Both of the above procedures introduce more complicated operators than appear in Eqs. (\ref{Eq3})-(\ref{Eq5}).  The more complicated operators may be eliminated from the direct Pauli reduction by noting the following:
\begin{eqnarray}
\label{Eq17}
\vec{\sigma }\cdot \vec{p} = \frac{1}{r^{2}} \left(\vec{\sigma }\cdot \vec{r}\right)  \left(\vec{\sigma }\cdot \vec{r}\right)  \left(\vec{\sigma} \cdot \vec{p}\right)\nonumber\\
= \frac{1}{r^{2}} \left(\vec{\sigma } \cdot \vec{r}\right)  \left[\vec{r} \cdot \vec{p} +i \vec{\sigma } \cdot \left(\vec{r}\times\vec{p}\right) \right] \nonumber\\
= \left(\vec{\sigma } \cdot \vec{r}\right)  \left[ -\frac{i}{r}\frac{\partial}{\partial r}- \frac{i}{r} + \frac{i}{r} \left(1+ \vec{\sigma } \cdot \vec{L} \right)\right] \nonumber\\
= \left(\vec{\sigma } \cdot \vec{r}\right) \left[ -\frac{i}{r} \frac{\partial}{\partial r} r+ \frac{i}{r}\left(1+ \vec{\sigma } \cdot \vec{L} \right)\right]  .
\end{eqnarray}
This expression operates on $\eta_{U}$ giving
\begin{eqnarray}
\label{Eq18}
\eta_{L}=i\frac{G\left(r\right)}{r} \phi_{-\kappa}\rightarrow \frac{\vec{\sigma}\cdot \vec{p}}{E+M} \frac{F\left(r\right)}{r} \phi_{\kappa} \nonumber\\
=\frac{1}{E+M} \left(\frac{i}{r} \frac{\partial F\left(r\right)}{\partial r}+i\kappa\frac{F\left(r\right)}{r^{2}}  \right)  \phi_{-\kappa} .
\end{eqnarray}
Therefore, one can just replace $G(r)$ with $\tilde{G} = (F^{\prime} + \kappa F/r)/(E + M)$ and simplify by letting $E + M =  2M_{N}$.  This procedure has two advantages.  First one need not calculate any new operators in order to include the second-order terms; and second, one can see which terms arise from the lower component. 

At this point it appears that one can include relativistic effects to order $1/M^{2}_{N}$ in non-relativistic shell model calculations, and these terms should give significant contributions at high energy and momentum transfer.  However, if one does look at which terms arise from the lower component, one notes that every term in Eqs.(\ref{Eq4}) and  (\ref{Eq5}) comes from the lower component except those that depend on spin.  This should raise some concern when calculating transverse responses. To test the extent of this concern, one can look at transverse form factors in the relativistic continuum shell model \cite{GH12} in the Tamm-Dancoff approximation. This model provides continuum solutions in which binary breakup channels satisfy a relative Dirac equation.  These calculations also provide bound states whose particle wave functions are expansions of Dirac oscillators, much like a Hartree-Fock calculation.  In Ref. \cite{GH12} it was determined that even the simple $\sigma + \omega + \rho$ exchange with quantum hydrodynamics (QHD) coupling constants \cite{SW86} provided reasonable agreement with experimentally determined single-particle energies and the experimental $^{15}$N$(p,p)^{15}$N cross section at 39.84 MeV.

\begin{figure}[ht]
\includegraphics[width=8cm]{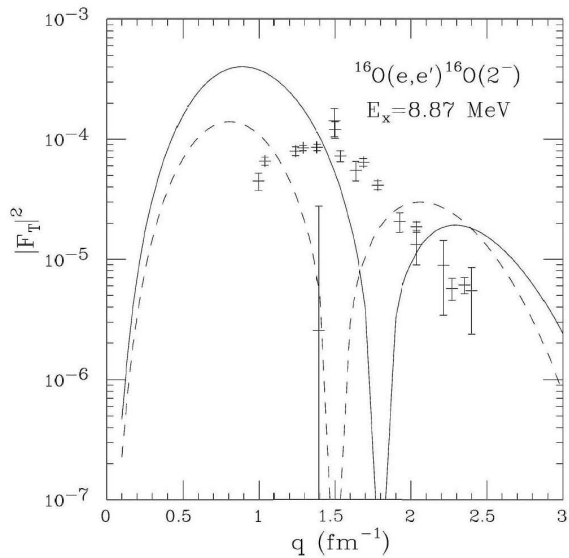}
\caption{\label{Fig1} Transverse form factor for $^{16}$O$(e,e^{\prime})^{16}$O$\left(2^{-}\right)$.  Solid line uses lower component; dashed line uses approximation in Eq. (\ref{Eq13}) for lower component. The data is from Ref. \cite{Hw84} as analyzed in Ref. \cite{F85}.}
\end{figure}
Calculations are performed for electro-excitation of the lowest $2^{-}$ and $3^{-}$ states in $^{16}$O with nucleon form factors evaluated at $q_{\mu}=0$.  Because the energy transfer is small, only terms of order $1/M_{N}$ have significant contributions. The calculated transverse form factor for the abnormal parity, $2^{-}$ state is shown in Fig. \ref{Fig1}.  The data is from Ref. \cite{Hw84} as analyzed in Ref. \cite{F85}.  The solid line is the complete calculation with Eqs. (\ref{Eq9})-(\ref{Eq10}), while the dashed line results from making the substitution in Eq. (\ref{Eq16}) for the lower component.   While neither calculation fits the data, one sees a very large difference between the two calculations.  The difference is almost entirely in the first-order terms.  Therefore, the substitution of Eq. (\ref{Eq16}), and hence the commonly used expressions in Eqs. (\ref{Eq4}) and (\ref{Eq5})  cannot account for the large relativistic effect.

\begin{figure}[ht]
\includegraphics[width=8cm]{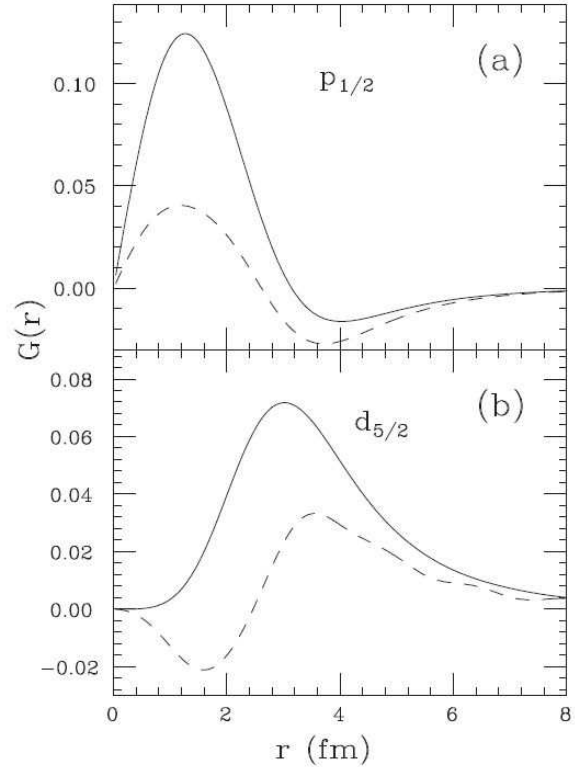}
\caption{\label{Fig2} Lower components of $d_{5/2^{-}}p^{-1}_{1/2}$ proton configuration for the $2^{-}$, $T=0$ state.  Solid line is the lower component; dashed line is approximation of Eq. (\ref{Eq16}).}
\end{figure}
\begin{figure}[ht]
\includegraphics[width=8cm]{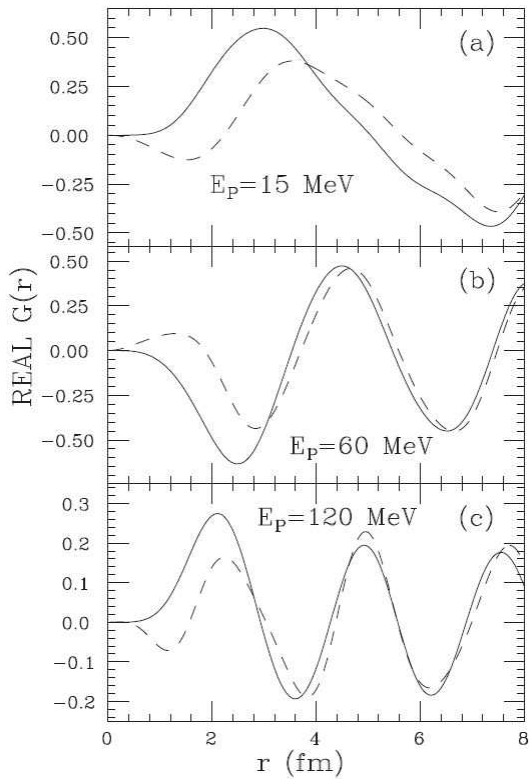}
\caption{\label{Fig3} Real part of the lower component for $d_{5/2}$ proton in the continuum.  Solid lines are the lower component; dashed lines are approximation of Eq. (\ref{Eq16}).}
\end{figure}
The source of this error is demonstrated in Fig. \ref{Fig2} where the lower component and Eq. (\ref{Eq16}) are plotted.  For the dominant $d_{5/2^{-}}p^{-1}_{1/2}$ proton component, the two are very different.  One might expect this component to show a large difference since relativistic effects were shown in Refs. \cite{U99,U98}  to be larger for excitation from the $p_{1/2}$ shell than the $p_{3/2}$ shell, and in general, larger for the $j=l-1/2 $  than the for the $j=l+1/2$  bound state.  The difference for the $d_{5/2}$ particle state decreases as one goes to the continuum and decreases as the energy is increased as shown in Fig. \ref{Fig3}.  By 120 MeV the difference is small, so one expects the error will be less for knockout reactions at high energy transfer, which is also where one expects the second order terms to contribute.  However, the initial $p_{1/2}$ state will still be different.

\begin{figure}[ht]
\includegraphics[width=8cm]{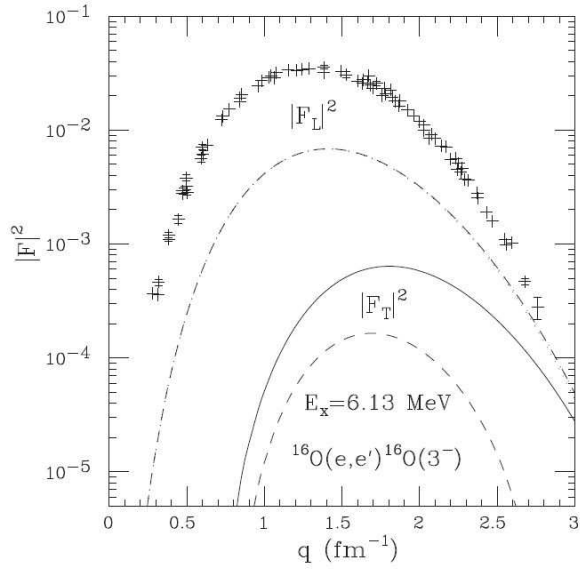}
\caption{\label{Fig4} Form factors for the $^{16}$O$(3^{–})$ state.  Solid line is transverse form factor using lower component; dashed line uses approximation in Eq. (\ref{Eq16}) for lower component. Dot-dashed line is the longitudinal form factor using lower component; dotted line uses approximation in Eq. (\ref{Eq16}).  Longitudinal data is from Ref. \cite{B86}.}
\end{figure}
The transverse form factor for the $3^{-}$ state is shown in Fig. \ref{Fig4}, with and without the substitution in Eq. (\ref{Eq16}), and the same large difference is obtained.  Also shown is the longitudinal form factor with (dotted line) and without (dot-dashed line) the substitution and the data of Ref. \cite{B86}.  The two calculations are nearly indistinguishable because the substitution for the longitudinal form factor contributes only to the $1/M^{2}_{N}$ terms.  Therefore, longitudinal form factors show little dependence on relativistic effects, but the transverse for factors have an inherent uncertainty, even at low energy transfer.
 
In conclusion, it was shown that corrections to multipole operators of order $1/M^{2}_{N}$ can be easily calculated if one has a code that can calculate the first-order terms.  The second order terms should contribute at high energy and momentum transfer.  For bound states the second order terms are negligible, but calculations for transverse form factors have substantial uncertainties due to relativistic effects.

\section*{Acknowledgments}
This work was supported by the National Science Foundation under Grant No. PHY-0855339.

\end{document}